\documentclass[prd,twocolumn,showpacs,preprintnumbers,amsmath,amssymb,superscriptaddress,floatfix,nofootinbib]{revtex4}

\usepackage{graphicx}
\usepackage{amsmath}
\usepackage{amsfonts}
\usepackage{amssymb}
\usepackage{color}
\usepackage{multirow}
\usepackage[colorlinks, citecolor=blue,anchorcolor=red,menucolor=red, linkcolor=red,filecolor=red,runcolor=red,urlcolor=blue,frenchlinks=red]{hyperref}

\begin{document}

\title{Role of $Y(4630)$ in the $p\bar{p}\rightarrow\Lambda_c\bar{\Lambda}_c$ reaction near threshold}

\author{Yan-Yan Wang}
\affiliation{Department of Physics, Zhengzhou University, Zhengzhou, Henan 450001, China}

\author{Qi-Fang L\"{u}} \email{lvqifang@ihep.ac.cn}
\affiliation{Institute of High Energy Physics, Chinese Academy of Sciences, Beijing 100049, China}

\author{En Wang} \email{wangen@zzu.edu.cn}
\affiliation{Department of Physics, Zhengzhou University, Zhengzhou, Henan 450001, China}

\author{De-Min Li} \email{lidm@zzu.edu.cn}
\affiliation{Department of Physics, Zhengzhou University, Zhengzhou, Henan 450001, China}

\begin{abstract}

We investigate the charmed baryon production reaction $p\bar{p}\rightarrow\Lambda_c\bar{\Lambda}_c$ in an effective Lagrangian approach. Besides the $t$-channel $D^0$ and $D^{*0}$ mesons exchanges, the $s$-channel $Y(4630)$ meson exchange is taken into account. For the total cross sections, the $D^0$ and $D^{*0}$ mesons provide minor background contributions, while the $Y(4630)$ state gives a clear peak structure with the magnitude of 10 $\mu$b at center of mass energy 4.63 GeV. Basing on the results, we suggest that the reaction of $p\bar{p}\rightarrow\Lambda_c\bar{\Lambda}_c$ can be used to search for the $1^{--}$ charmonium-like $Y(4630)$ state, and our predictions can be tested in future by the $\rm{\bar PANDA}$ facility.
\end{abstract}
\date{\today}
 \maketitle

\section{Introduction}{\label{Introduction}}

A new charmonium-like  $Y(4630)$, $J^{PC}=1^{--}$, was firstly reported by the Belle collaboration in the exclusive $e^+e^- \to \Lambda_c \bar \Lambda_c$ process, and its mass and width are $4634^{+8+5}_{-7-8}~\rm{MeV}$ and $92^{+40+10}_{-24-21}~\rm{MeV}$, respectively~\cite{Pakhlova:2008vn}. After its discovery, various interpretations, such as conventional charmonium state~\cite{Badalian:2008dv,Segovia:2008ta}, tetraquark state~\cite{Maiani:2014aja,Cotugno:2009ys,Brodsky:2014xia}, $\Lambda_c \bar \Lambda_c$ baryonium~\cite{Lee:2011rka,Chen:2011cta} and threshold effect~\cite{vanBeveren:2008rt}, were performed. The mechanism of $Y(4630)$ enhancement in $\Lambda_c \bar \Lambda_c$ electroproduction was also studied~\cite{Simonov:2011jc}. Recently, a series of investigations on the strong decay behaviors were proposed, which intended to reveal the inner structure of the $Y(4630)$~\cite{Liu:2016sip,Liu:2016nbm,Guo:2016iej}.

Above the $\Lambda_c \bar \Lambda_c$ threshold, another $1^{--}$ resonance $Y(4660)$, which mass and width are consistent within errors with the state $X(4630)$, was  observed in the initial state radiation process $e^+e^- \to \gamma_{\rm ISR} \pi^+ \pi^- \psi(2S)$ by the Belle collaboration~\cite{Wang:2007ea}, and confirmed by the BaBar collaboration after a long debate~\cite{Lees:2012pv}. The explanations include conventional $c\bar c$ state~\cite{Ding:2007rg,Li:2009zu}, $\psi(2S) f_0(980)$ molecule~\cite{Guo:2008zg,Guo:2009id,Albuquerque:2011ix}, hadro-charmonium~\cite{Dubynskiy:2008mq}, tetraquark state~\cite{Maiani:2014aja,Ebert:2008kb,Chen:2010ze} and baryonium~\cite{Qiao:2007ce}. Although the $Y(4630)$ and $Y(4660)$ were observed in different processes, the similar masses and widths suggest that they could be the same state, which is also discussed in many theoretical works~\cite{Cotugno:2009ys,Bugg:2008sk,Guo:2010tk}. Other related studies are also performed~\cite{Cleven:2015era,Chen:2013wca,Chen:2013axa,Chen:2015bma}, and a comprehensive review can be found in Ref.~\cite{Chen:2016qju}. In the present work, we adopt the wildly accepted opinion that the $Y(4630)$ and $Y(4660)$ are regarded as the same state.

The theoretical works of the $Y(4630)$ mainly focus on the mass and decay width, and the production experiment is only limited in $e^+e^-$ collision.
In addition, the production of charmed baryon states in the $p\bar{p}$ collisions has been investigated within many theoretical models, such as the quark-diquark picture, a handbag approach, a quark-gluon string model which based on Regge asymptotics, meson-exchange model, and single-channel effective Lagrangian model~\cite{Kroll:1988cd,Kaidalov:1994mda,Titov:2008yf,Goritschnig:2009sq,Haidenbauer:2009ad,Haidenbauer:2010nx,Khodjamirian:2011sp,Shyam:2014dia}.

Taking into account that the branching ratio of $Y(4630) \to p \bar p$ partial decay process has several percent by the prediction of Ref.~\cite{Guo:2016iej}, we suggest to search the $Y(4630)$ state in the reaction of $p \bar p \to \Lambda_c \bar \Lambda_c$, where the intermediate state $Y(4630)$ is wished to play an important role.
In the present work, we will study the reaction of  $p \bar p \to \Lambda_c \bar \Lambda_c$ within the effective Lagrangian approach by taking into account the  $t$-channel $D$ and $D^*$ mesons exchanges, the $s$-channel $Y(4630)$ contribution, and predict the  total and differential cross sections of the reaction $p \bar p \to \Lambda_c \bar \Lambda_c$. Our predictions can be tested by the  $\rm{\bar PANDA}$ facility, which has a maximum beam momenta 15 GeV of antiproton with high luminosity~\cite{Wiedner:2011mf}, and is enough to produce the $Y(4630)$ state.

This paper is organized as follows. We present the formalism and ingredients of the effective Lagrangian approach in Sec.~\ref{FORMALISM AND INGREDIENTS}. The numerical results of total and differential cross sections and discussions are shown in Sec.~\ref{NUMERICAL RESULTS AND DISCUSSIONS}. Finally, a short summary is given in the last section.

\section{FORMALISM AND INGREDIENTS}{\label{FORMALISM AND INGREDIENTS}}

\begin{figure}[htbp]
\includegraphics[scale=1.0]{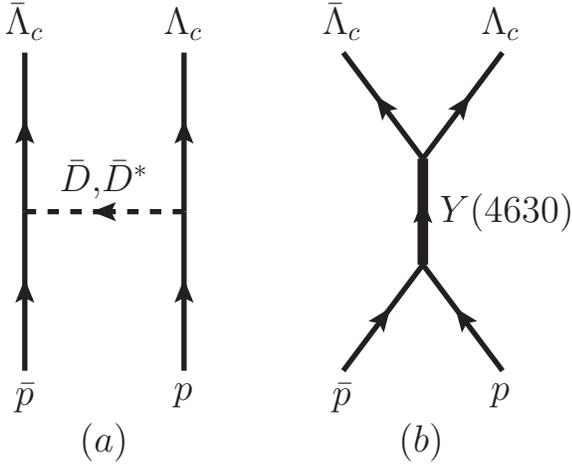}
\vspace{0.0cm} \caption{Feynman diagrams for the $p\bar{p}\rightarrow\Lambda_c\bar{\Lambda}_c$ reaction. a) the $t$-channel $D$ and $D^*$ meson exchanges, b) the $s$-channel $Y(4630)$ term. }
\label{feyn}
\end{figure}

For the process $p\bar{p}\rightarrow\Lambda_c\bar{\Lambda}_c$, we will take into account the basic tree level Feynman diagrams, depicted in Fig.~\ref{feyn},  which include the $t$-channel $D$ and $D^*$ meson exchanges, the $s$-channel $Y(4630)$ term.

The relevant effective Lagrangians of the vertexes in Fig.~\ref{feyn} can be written as~\cite{Xie:2015zga,Guo:2016iej,He:2011jp},
\begin{equation}
\mathcal{L}_{\Lambda_c p D} = ig_{\Lambda_c p D} \bar{\Lambda}_c \gamma_5 p D + {\rm H.c.},
\end{equation}
\begin{equation}
\mathcal{L}_{\Lambda_c p D^*} = g_{\Lambda_c p D^*} \bar{\Lambda}_c \gamma^{\mu} p D^*_{\mu} + {\rm H.c.},
\end{equation}
\begin{equation}
\mathcal{L}_{Y \Lambda_c \bar \Lambda_c} = g_{Y \Lambda_c \bar \Lambda_c} Y_{\mu} \bar \Lambda_c \gamma^{\mu} \Lambda_c,
\end{equation}
\begin{equation}
\mathcal{L}_{Y p \bar p} = g_{Y p \bar p} Y_{\mu} \bar{p} \gamma^{\mu} p,
\end{equation}
where the coupling constants of $\Lambda_c p D$ and $\Lambda_c p D^*$ interactions are taken as $g_{\Lambda_c p D}=10.7^{+5.3}_{-4.3}$ and $g_{\Lambda_c p D^*}=-5.8^{+2.1}_{-2.5}$~\cite{Khodjamirian:2011sp,Khodjamirian:2011jp}. The couplings of $Y \Lambda_c \bar \Lambda_c$ and $Y p \bar p$ can be obtained from their partial decay widths. In Ref.~\cite{Guo:2016iej}, assuming the $Y(4630) \to \Lambda_c \bar \Lambda_c$ is dominant decay, the branching ratios of $p \bar p$ final state are predicted to be $0.037$ and $0.062$ for different cut off values. In this work, we adopt the same assumption of the $\Lambda_c \bar \Lambda_c$ being the dominant decay channel, and take the $p\bar p$ decay ratio being $1\%$. With the effective Lagrangians above, the partial decay widths can be expressed as,
\begin{equation}
\Gamma(Y(4630) \to \Lambda_c \bar \Lambda_c) = \frac{g_{Y \Lambda_c \bar \Lambda_c}^2(m_Y^2 + 2 m_{\Lambda_c}^2)|\vec p_{\Lambda_c}^{\,\rm cm}|}{6\pi m_Y^2},
\end{equation}
\begin{equation}
\Gamma(Y(4630) \to p \bar p) = \frac{g_{Y p \bar p}^2(m_Y^2 + 2 m_p^2)|\vec p_p^{\, \rm cm}|}{6\pi m_Y^2},
\end{equation}
where $m_Y$, $m_{\Lambda_c}$ and $m_p$ are the masses of $Y(4630)$, $\Lambda_c$/$\bar {\Lambda}_c$ and (anti)proton~\cite{Agashe:2014kda}, respectively, and $\vec p_{\Lambda_c}^{\,\rm cm}$ ($\vec p_p^{\, \rm cm}$) is the 3-momentum of the initial proton (final $\Lambda_c$) in the rest frame of $p\bar{p}$ ($\Lambda_c\bar{\Lambda}_c$). With the experimental data of the total decay width  $\Gamma_Y=92$ MeV\cite{Pakhlova:2008vn}, we can obtain $g_{Y \Lambda_c \bar \Lambda_c} = 1.78$, and $g_{Y p \bar p} = 0.087$.

Since the hadrons are not point-like particles, the form factors are needed to describe the off-shell effects. We adopt here the monopole form factor used in many previous works for the $t$-channel $D$ and $D^*$ interaction vertices:
\begin{equation}
\mathcal{F}(q^2, m^2) = \frac{\Lambda^2-m^2}{\Lambda^2-q^2},
\end{equation}
where $q$, $m$ and $\Lambda$ are the the four-momentum, mass, and cut-off parameter for the exchanged mesons, respectively. The cut-off parameter $\Lambda$ can be parametrized as~\cite{Lin:2014jza}
\begin{equation}
\Lambda = m+ \alpha\Lambda_{\rm QCD},
\end{equation}
with $\Lambda_{\rm QCD} = 220$ MeV, and the dimensionless parameter $\alpha$ is of order unity. The $\alpha$ mainly varies in the range of $0.5 \sim 1.5$ in literature~\cite{Lin:2014jza,Chen:2012nva,Chen:2014ccr}, and the $D$ and $D^*$ meson, the large value of $\alpha = 1.5$ is employed in present work. Indeed, the value of $\alpha$ does not affect the signal of the $Y(4630)$ in the reaction $p\bar{p}\rightarrow\Lambda_c\bar{\Lambda}_c$. The form factor for $s$-channel $Y(4630)$ state is taken in the form advocated in Refs.~\cite{Feuster:1998pq,Penner:2002ma,Shklyar:2005xg,Kim:2011rm,Wang:2015jsa,Lu:2015fva,Shyam:2015hqa},
\begin{equation}
F_Y(q^2,m^2) = \frac{\Lambda_Y^4}{\Lambda_Y^4+(q^2-m_Y^2)^2},
\end{equation}
where the cut-off parameters $\Lambda_Y=500$ MeV for the $Y(4630)$ state is used~\cite{Kim:2011rm,Wang:2015jsa,Lu:2015fva}.

Then, according to the Feynman rules, the scattering amplitudes for the $p\bar{p}\rightarrow\Lambda_c\bar{\Lambda}_c$ reaction can be obtained straightforwardly with the above effective Lagrangians,
\begin{eqnarray}
\mathcal{M}_{D} &=& g_{\Lambda_c p D}^2 \mathcal{F}^2(q^2_D,m_D^2)\bar{\upsilon}(p_1,s_1)\gamma_5\upsilon(p_3,s_3)\nonumber
\\&&G_D\bar{u}(p_4,s_4)\gamma_5u(p_2,s_2),
\end{eqnarray}
\begin{eqnarray}
\mathcal{M}_{D^*} &=& -g_{\Lambda_c p D^*}^2\mathcal{F}^2(q^2_{D^*},m_{D^*}^2)\bar{\upsilon}(p_1,s_1)\gamma_{\mu}\upsilon(p_3,s_3)\nonumber
\\&&G_{D^*}^{\mu\nu}\bar{u}(p_4,s_4)\gamma_{\nu}u(p_2,s_2),
\end{eqnarray}
\begin{eqnarray}
\mathcal{M}_Y &=& -g_{Y \Lambda_c \bar \Lambda_c} g_{Y p \bar p} F_Y(q_Y^2,m_Y^2)\bar{\upsilon}(p_1,s_1)\gamma_{\mu}\upsilon(p_2,s_2)\nonumber
\\&&G_Y^{\mu\nu}\bar{u}(p_4,s_4)\gamma_{\nu}u(p_3,s_3),
\end{eqnarray}
where $s_i(i=1,2,3,4)$ and $p_i(i=1,2,3,4)$ represent the spin projection and four-momentum of the initial or final states, respectively; $q_{D^{(*)}} = p_3 - p_1$ is the momentum of $D^0$ ($D^{*0}$) state. $q_Y = p_1 + p_2$ is the momentum of the $Y(4630)$ state. $G_D$, $G_{D^*}$ and $G_Y$ are the propagators for the $D$, $D^*$, and $Y(4630)$ states.

The $D$ meson propagator can be written as
\begin{equation}
G_{D}= \frac{i}{q^2-m^2_D},
\end{equation}
and the one of vector meson $D^*$ is
\begin{equation}
G_{D^*}^{\mu\nu}= -i\frac{g^{\mu\nu}-q^{\mu}q^{\nu}/m_{D^*}^2}{q^2-m_{D^*}^2}.
\end{equation}
The propagator for $Y(4630)$ $1^{--}$ state can be written as,
\begin{equation}
G_Y =  -i\frac{g^{\mu\nu}-q^{\mu}q^{\nu}/m_Y^2}{q^2-m_Y^2+im_Y \Gamma_Y},
\end{equation}
where $\Gamma_Y=92~\rm{MeV}$ is the total width of the $Y(4630)$ meson.

The total amplitude for the process $p\bar{p}\rightarrow\Lambda_c\bar{\Lambda}_c$ are the coherent sum of $\mathcal{M}_D$, $\mathcal{M}_{D^*}$, and $\mathcal{M}_Y$,
\begin{eqnarray}
\mathcal{M} &=& \mathcal{M}_D+\mathcal{M}_{D^*}+\mathcal{M}_Y.
\end{eqnarray}
The differential cross section can be easily given as,
\begin{eqnarray}
\frac{{\rm d}\,\sigma}{{\rm d\,cos}\theta} = \frac{1}{32\pi s}\frac{|\vec{p}^{\rm ~c.m.}_3|}{|\vec{p}^{\rm ~c.m.}_1|}\left(\frac{1}{4}\sum_{s_1,s_2,s_3,s_4}|\mathcal{M}|^2\right)
\end{eqnarray}
where $s$ is the invariant mass square of the $p\bar{p}$ system. $\theta$ denotes the angle of the outgoing baryon $\Lambda_c$ relative to the beam direction in the c.m. frame, while $\vec{p}^{\rm ~c.m.}_1$ and $\vec{p}^{\rm ~c.m.}_3$ are the 3-momentum of the initial $p$ and final $\Lambda_c$ in the c.m. frame.

\section{NUMERICAL RESULTS AND DISCUSSIONS}{\label{NUMERICAL RESULTS AND DISCUSSIONS}}

In this section, we will present our results for the $p\bar{p}\rightarrow\Lambda_c\bar{\Lambda}_c$ reaction within the effective Lagrangian model.

\begin{figure}[!htbp]
\includegraphics[scale=0.8]{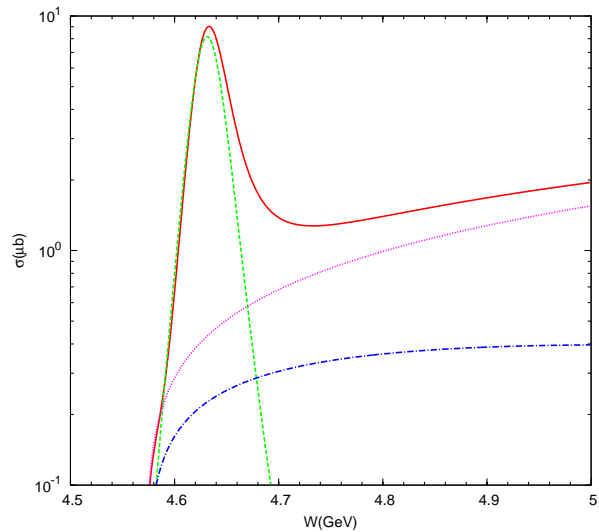}
\vspace{0.0cm} \caption{(online color) Cross section of the $p\bar{p}\rightarrow\Lambda_c\bar{\Lambda}_c$ reaction as a function of $p\bar{p}$ invariant mass energies from threshold up to $5~\rm{GeV}$. The green dashed, blue dash-dotted, and pink dotted lines represent the contributions of the $s$-channel $Y(4630)$ terms, $t$-channel $D$ and $D^*$ mesons exchanges, respectively. The red solid line stands for the total cross section.}
\label{tcs}
\end{figure}

In Fig.~\ref{tcs}, we show the total cross sections for the $p\bar{p}\rightarrow\Lambda_c\bar{\Lambda}_c$ reaction from threshold up to 5 GeV of c.m.~energy. In the figure, the green dashed,  blue dash-dotted, and pink  dotted lines represent the contributions of the $s$-channel $Y(4630)$ terms, $t$-channel $D$ and $D^*$ mesons exchanges, respectively. The red solid line stands for the total cross section. The $t$-channel $D^*$ meson exchange provides a larger contribution than $D$ meson exchange, which are consistent with the result of Ref.~\cite{Shyam:2014dia}. Indeed, the Ref.~\cite{Shyam:2014dia} gives a larger total cross section than our work, which
could be the reason that Ref.~\cite{Shyam:2014dia} takes into account the contribution of the tensor coupling term in the $t$-channel $D^*$ meson exchange (see Fig.~3 of Ref.~\cite{Shyam:2014dia}).  Yet, since our purpose is to see if the $Y(4630)$ state plays an important role in this reaction,
and if the signal of the $Y(4630)$ state can be seen in the $\rm{\bar PANDA}$ facility, it is no problem to neglect the contribution of the tensor term in the $t$-channel $D^*$ meson exchange.  In Fig.~\ref{tcs}, by including the contribution of $s$-channel, the total cross section has a clear peak structure at the  c.m.~energy 4630 MeV. The magnitude of the peak is of order $10~\mu$b, which can be enough measured in future $\rm{\bar PANDA}$ facility with high luminosity, thus we can safely neglect the tensor term in the $t$-channel $D^*$ meson exchange.

It is interesting to note that for the $Y(4630)$ state, a relatively small cut off value is used, which seems more suitable for heavy hadron production processes~\cite{Kim:2011rm,Wang:2015jsa,Lu:2015fva}. As this value increases, the contributions of $Y(4630)$ will become larger and the peak structure is also clear. In fact, the form factor is approximate equal to unity near the $Y(4630)$ resonance region, since the $s-M_Y^2 \sim 0$.
Thus, whatever value of the cut-off for the $Y(4630)$ state, its signal is always clear, which is one of the reasons why we suggest to search the $Y(4630)$ state in the $p\bar{p}\rightarrow\Lambda_c\bar{\Lambda}_c$ reaction.

\begin{figure}[!htbp]
\includegraphics[scale=0.85]{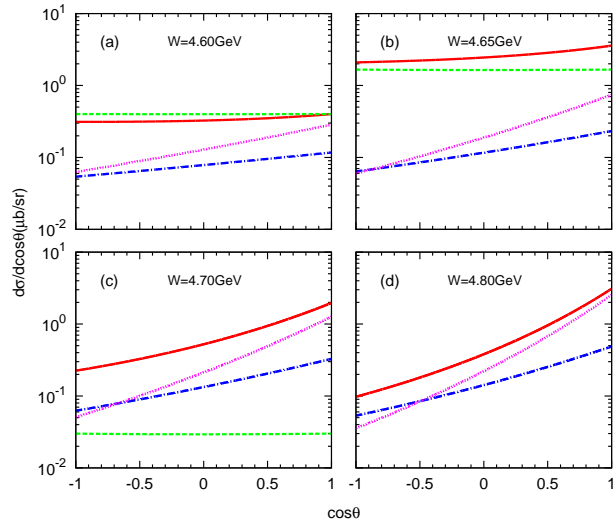}
\vspace{0.0cm} \caption{Differential cross section of the $p\bar{p}\rightarrow\Lambda_c\bar{\Lambda}_c$ reaction as a function of cos$\theta$. The explanations of the curves are same as that of Fig.~\ref{tcs}.}
\label{dcs}
\end{figure}

In addition to the total cross sections, we also show the differential cross sections of $p\bar{p}\rightarrow\Lambda_c\bar{\Lambda}_c$ reaction for different center-of-mass energy $\sqrt{s}$ in Fig.~\ref{dcs}. In the energy region from 4.6 to 4.8 GeV, the contributions of the $t$-channel $D$ and $D^*$ exchanges have the forward distributions, while the ones of the $Y(4630)$ are much flat. Close to the energy of 4.63 GeV, the $Y(4630)$ plays the dominate role, and its contribution can be ignored above 4.7~GeV.
Thus, the differential cross sections of our total model are much flatter around the 4.63 GeV than that above 4.7 GeV, which will give a signature of the resonance state $Y(4630)$.

\section{Summary}

Considering the $J^{PC}=1^{--}$ charmonium-like state $Y(4630)$ have a sizeable coupling to the $p\bar{p}$ according the prediction of Ref.\cite{Guo:2016iej}, we suggest to search the $Y(4630)$ state in the reaction of $p \bar p \to \Lambda_c \bar \Lambda_c$ in this work.

Within an effective Lagrangian approach, we have phenomenologically investigated the $p\bar{p}\rightarrow\Lambda_c\bar{\Lambda}_c$ reaction. Besides the background contributions of the $t$-channel $D$ and $D^*$ mesons exchanges, we have also included the $s$-channel $Y(4630)$ contribution. We have presented the total cross sections and the differential cross sections for this process. Our results show that close to the threshold of $p\bar{p}$ collision, the $Y(4630)$ state plays an important role, comparing to the background terms of the $t$-channel $D$ and $D^*$ mesons exchanges, and a clear bump structure with the magnitude of 10 $\mu$b appears.  Our predictions can be tested in future by the  $\rm{\bar PANDA}$ facility, which has a maximum beam momenta 15 GeV of antiproton with high luminosity~\cite{Wiedner:2011mf}, and is capable of producing the $Y(4630)$ state.

\section*{Acknowledgements}
We would like to thank Yu-Bing Dong and Guan-Nan Li for valuable discussions. This work is partly supported by the National Natural Science Foundation of China under Grant No. 11505158, the China Postdoctoral Science Foundation (No. 2015M582197), and the Postdoctoral Research Sponsorship in Hean Province (No. 2015023).


\begin{thebibliography}{99}

\bibitem{Pakhlova:2008vn}
  G.~Pakhlova {\it et al.} [Belle Collaboration],
  Phys.\ Rev.\ Lett.\  {\bf 101}, 172001 (2008).

\bibitem{Badalian:2008dv}
  A.~M.~Badalian, B.~L.~G.~Bakker and I.~V.~Danilkin,
  Phys.\ Atom.\ Nucl.\  {\bf 72}, 638 (2009).

\bibitem{Segovia:2008ta}
  J.~Segovia, D.~R.~Entem and F.~Fernandez,
  arXiv:0810.2875.

\bibitem{Maiani:2014aja}
  L.~Maiani, F.~Piccinini, A.~D.~Polosa and V.~Riquer,
  Phys.\ Rev.\ D {\bf 89}, 114010 (2014).


\bibitem{Cotugno:2009ys}
  G.~Cotugno, R.~Faccini, A.~D.~Polosa and C.~Sabelli,
  Phys.\ Rev.\ Lett.\  {\bf 104}, 132005 (2010).


\bibitem{Brodsky:2014xia}
  S.~J.~Brodsky, D.~S.~Hwang and R.~F.~Lebed,
  Phys.\ Rev.\ Lett.\  {\bf 113}, no. 11, 112001 (2014).


\bibitem{Lee:2011rka}
  N.~Lee, Z.~G.~Luo, X.~L.~Chen and S.~L.~Zhu,
  Phys.\ Rev.\ D {\bf 84}, 014031 (2011).

\bibitem{Chen:2011cta}
  Y.~D.~Chen and C.~F.~Qiao,
  Phys.\ Rev.\ D {\bf 85}, 034034 (2012)
.


\bibitem{vanBeveren:2008rt}
  E.~van Beveren, X.~Liu, R.~Coimbra and G.~Rupp,
  Europhys.\ Lett.\  {\bf 85}, 61002 (2009).

\bibitem{Simonov:2011jc}
  Y.~A.~Simonov,
  Phys.\ Rev.\ D {\bf 85}, 105025 (2012).

\bibitem{Liu:2016sip}
  X.~Liu, H.~W.~Ke, X.~Liu and X.~Q.~Li,
  arXiv:1601.00762.

\bibitem{Liu:2016nbm}
  X.~Liu, H.~W.~Ke, X.~Liu and X.~Q.~Li,
  arXiv:1602.00226.

\bibitem{Guo:2016iej}
  X.~D.~Guo, D.~Y.~Chen, H.~W.~Ke, X.~Liu and X.~Q.~Li,
  Phys.\ Rev.\ D {\bf 93}, no. 5, 054009 (2016).




\bibitem{Wang:2007ea}
  X.~L.~Wang {\it et al.} [Belle Collaboration],
  Phys.\ Rev.\ Lett.\  {\bf 99}, 142002 (2007).

\bibitem{Lees:2012pv}
  J.~P.~Lees {\it et al.} [BaBar Collaboration],
  Phys.\ Rev.\ D {\bf 89}, no. 11, 111103 (2014).

\bibitem{Ding:2007rg}
  G.~J.~Ding, J.~J.~Zhu and M.~L.~Yan,
  Phys.\ Rev.\ D {\bf 77}, 014033 (2008).

\bibitem{Li:2009zu}
  B.~Q.~Li and K.~T.~Chao,
  Phys.\ Rev.\ D {\bf 79}, 094004 (2009).

\bibitem{Guo:2008zg}
  F.~K.~Guo, C.~Hanhart and U.~G.~Meissner,
  Phys.\ Lett.\ B {\bf 665}, 26 (2008).

\bibitem{Guo:2009id}
  F.~K.~Guo, C.~Hanhart and U.~G.~Meissner,
  Phys.\ Rev.\ Lett.\  {\bf 102}, 242004 (2009).

\bibitem{Albuquerque:2011ix}
  R.~M.~Albuquerque, M.~Nielsen and R.~R.~da Silva,
  Phys.\ Rev.\ D {\bf 84}, 116004 (2011).


\bibitem{Dubynskiy:2008mq}
  S.~Dubynskiy and M.~B.~Voloshin,
  Phys.\ Lett.\ B {\bf 666}, 344 (2008).

\bibitem{Ebert:2008kb}
  D.~Ebert, R.~N.~Faustov and V.~O.~Galkin,
  Eur.\ Phys.\ J.\ C {\bf 58}, 399 (2008).


\bibitem{Chen:2010ze}
  W.~Chen and S.~L.~Zhu,
  Phys.\ Rev.\ D {\bf 83}, 034010 (2011).

\bibitem{Qiao:2007ce}
  C.~F.~Qiao,
  J.\ Phys.\ G {\bf 35}, 075008 (2008).

\bibitem{Bugg:2008sk}
  D.~V.~Bugg,
  J.\ Phys.\ G {\bf 36}, 075002 (2009).


\bibitem{Guo:2010tk}
  F.~K.~Guo, J.~Haidenbauer, C.~Hanhart and U.~G.~Meissner,
  Phys.\ Rev.\ D {\bf 82}, 094008 (2010).


\bibitem{Cleven:2015era}
  M.~Cleven, F.~K.~Guo, C.~Hanhart, Q.~Wang and Q.~Zhao,
  Phys.\ Rev.\ D {\bf 92}, no. 1, 014005 (2015).


\bibitem{Chen:2013wca}
  D.~Y.~Chen, X.~Liu and T.~Matsuki,
  Phys.\ Rev.\ Lett.\  {\bf 110}, no. 23, 232001 (2013).

\bibitem{Chen:2013axa}
  D.~Y.~Chen, X.~Liu and T.~Matsuki,
  J.\ Phys.\ G {\bf 42}, no. 1, 015002 (2015).

\bibitem{Chen:2015bma}
  D.~Y.~Chen, X.~Liu and T.~Matsuki,
  Phys.\ Rev.\ D {\bf 93}, no. 3, 034028 (2016).


\bibitem{Chen:2016qju}
  H.~X.~Chen, W.~Chen, X.~Liu and S.~L.~Zhu,
  arXiv:1601.02092.

\bibitem{Kroll:1988cd}
  P.~Kroll, B.~Quadder and W.~Schweiger,
  Nucl.\ Phys.\ B {\bf 316}, 373 (1989).

\bibitem{Kaidalov:1994mda}
  A.~B.~Kaidalov and P.~E.~Volkovitsky,
  Z.\ Phys.\ C {\bf 63}, 517 (1994).


\bibitem{Titov:2008yf}
  A.~I.~Titov and B.~Kampfer,
  Phys.\ Rev.\ C {\bf 78}, 025201 (2008).

\bibitem{Goritschnig:2009sq}
  A.~T.~Goritschnig, P.~Kroll and W.~Schweiger,
  Eur.\ Phys.\ J.\ A {\bf 42}, 43 (2009).

\bibitem{Haidenbauer:2009ad}
  J.~Haidenbauer and G.~Krein,
  Phys.\ Lett.\ B {\bf 687}, 314 (2010).

\bibitem{Haidenbauer:2010nx}
  J.~Haidenbauer and G.~Krein,
  Few Body Syst.\  {\bf 50}, 183 (2011).

\bibitem{Khodjamirian:2011sp}
  A.~Khodjamirian, C.~Klein, T.~Mannel and Y.~M.~Wang,
  Eur.\ Phys.\ J.\ A {\bf 48}, 31 (2012).


\bibitem{Shyam:2014dia}
  R.~Shyam and H.~Lenske,
  Phys.\ Rev.\ D {\bf 90}, no. 1, 014017 (2014).


\bibitem{Wiedner:2011mf}
  U.~Wiedner,
  Prog.\ Part.\ Nucl.\ Phys.\  {\bf 66}, 477 (2011).

\bibitem{Xie:2015zga}
  J.~J.~Xie, Y.~B.~Dong and X.~Cao,
  Phys.\ Rev.\ D {\bf 92}, no. 3, 034029 (2015).

\bibitem{He:2011jp}
  J.~He, Z.~Ouyang, X.~Liu and X.~Q.~Li,
  Phys.\ Rev.\ D {\bf 84}, 114010 (2011).


\bibitem{Khodjamirian:2011jp}
  A.~Khodjamirian, C.~Klein, T.~Mannel and Y.-M.~Wang,
  JHEP {\bf 1109}, 106 (2011).


\bibitem{Agashe:2014kda}
  K.~A.~Olive {\it et al.} (Particle Data Group Collaboration),
  Review of Particle Physics,
  Chin.\ Phys.\ C {\bf 38}, 090001 (2014).

\bibitem{Lin:2014jza}
  Q.~Y.~Lin, X.~Liu and H.~S.~Xu,
  Phys.\ Rev.\ D {\bf 90}, no. 1, 014014 (2014).

  
  
\bibitem{Chen:2012nva}
  D.~Y.~Chen, X.~Liu and T.~Matsuki,
  Phys.\ Rev.\ D {\bf 87}, no. 5, 054006 (2013).

\bibitem{Chen:2014ccr}
  D.~Y.~Chen, X.~Liu and T.~Matsuki,
  Phys.\ Rev.\ D {\bf 90}, no. 3, 034019 (2014).


\bibitem{Feuster:1998pq}
  T.~Feuster and U.~Mosel,
  Phys.\ Rev.\ C {\bf 58}, 457 (1998)

\bibitem{Penner:2002ma}
  G.~Penner and U.~Mosel,
  Phys.\ Rev.\ C {\bf 66}, 055211 (2002)

\bibitem{Shklyar:2005xg}
  V.~Shklyar, H.~Lenske and U.~Mosel,
  Phys.\ Rev.\ C {\bf 72}, 015210 (2005)


\bibitem{Kim:2011rm}
  S.~H.~Kim, S.~i.~Nam, Y.~Oh and H.~C.~Kim,
  Phys.\ Rev.\ D {\bf 84}, 114023 (2011).

\bibitem{Wang:2015jsa}
  Q.~Wang, X.~H.~Liu and Q.~Zhao,
  Phys.\ Rev.\ D {\bf 92}, 034022 (2015).


\bibitem{Lu:2015fva}
  Q.~F.~L\"{u}, X.~Y.~Wang, J.~J.~Xie, X.~R.~Chen and Y.~B.~Dong,
  Phys.\ Rev.\ D {\bf 93}, no. 3, 034009 (2016).

\bibitem{Shyam:2015hqa}
  R.~Shyam and H.~Lenske,
  Phys.\ Rev.\ D {\bf 93}, no. 3, 034016 (2016).






\end{thebibliography}
\end{document}